\documentstyle[aps,prc,preprint,epsf,psfig,hangcaption]{revtex}
\draft
\tightenlines

\newcommand{\K}{\enspace,}
\newcommand{\p}{\enspace.}
\newcommand{\be}{\begin{equation}}
\newcommand{\ee}{\end{equation}}
\newcommand{\bea}{\begin{eqnarray}}
\newcommand{\eea}{\end{eqnarray}}
\newcommand{\bc}{\begin{center}}
\newcommand{\ec}{\end{center}}

\newcommand{\btu}{\bigtriangleup}

\newcommand{\eps}{\varepsilon}

\newcommand{\lla}{\langle}
\newcommand{\gra}{\rangle}

\def \be {\begin{equation}}
\def \ee{\end{equation}}
\def \bea{\begin{eqnarray}}
\def \eea{\end{eqnarray}}
\def \l {\label}

\def \r {\ref}

\include{scrload}
\newfam\scrfam                                          
\global\font\twelvescr=rsfs10 scaled\magstep1%
\global\font\eightscr=rsfs7 scaled\magstep1%
\global\font\sixscr=rsfs5 scaled\magstep1%
\skewchar\twelvescr='177\skewchar\eightscr='177\skewchar\sixscr='177%
\textfont\scrfam=\twelvescr\scriptfont\scrfam=\eightscr
\scriptscriptfont\scrfam=\sixscr

\begin{document}

\title{Shear Viscosity Coefficient from Microscopic Models}
\address{}
\author{Azwinndini Muronga\footnote{Present address: {\it Institut f\"ur Theoretische
Physik, J.W. Goethe--Universit\"at, \\ \hspace*{3.3cm}D--60325 Frankfurt am
Main, Germany}}}
\address{ School of Physics and Astronomy, University of Minnesota, 
          Minneapolis, Minnesota 55455, USA}

\date{\today}
\maketitle

\begin{abstract}

The transport coefficient of shear viscosity is studied for a hadron matter
through microscopic transport model, the Ultra--relativistic Quantum Molecular
Dynamics (UrQMD), using the Green--Kubo formulas.     
Molecular--dynamical simulations are performed for a system of light mesons in
a box with periodic boundary conditions. Starting from an initial state
composed of $\pi,\,\eta\,,\omega\,,\rho\,,\phi$ with a uniform phase--space
distribution, the evolution takes place through elastic collisions, production
and annihilation.  The system approaches a stationary state of  mesons and
their resonances, which is characterized by common temperature.   After
equilibration, thermodynamic quantities such as the energy density, particle
density, and pressure are calculated.  From such an equilibrated state  the
shear viscosity coefficient is calculated from the fluctuations of stress tensor
around equilibrium using Green--Kubo relations. We do our simulations here at
zero net baryon density so that the equilibration times depend on the energy
density.  We do not include hadron strings as degrees of freedom so as to
maintain detailed balance. Hence we do not get the saturation of  temperature
but this leads to longer equilibration times.

\end{abstract}


\section{Introduction}

High energy heavy ion reactions are studied experimentally and theoretically to
obtain information about the properties of nuclear matter under the extreme
conditions of high densities and /or temperatures. One of the most important
aspects of studying nucleus-nucleus reactions at these extreme conditions is
the possibility that normal nuclear matter can undergo a phase transition into
a new state of matter, the quark--gluon plasma \cite{QMreviews}. In this state
the  degrees of freedom are partons (quarks and gluons). 

In this work we study only the the thermodynamic and transport properties of 
hadron matter. Hence the relevant degrees of freedom are hadrons. We study the
equilibration of the system in infinite hadron matter using UrQMD
\cite{Bass98}. We restrict ourselves to a a system that contains only meson
resonance degrees of freedom. The infinite hadron matter is modelled by
initializing the system by light mesons only. We fix baryon density and energy
density of the system in a cubic box and impose periodic boundary conditions.
We then propagate the system in time until we obtain equilibration.

The  equation of state and transport coefficients of hot, dense hadron gases
are quite important quantities in high energy nuclear physics. In the
ultra--relativistic heavy ion experiments at CERN and BNL, the final state of
interactions is dominated by hadrons and hence the observables are mainly
hadrons. Therefore knowledge of the equation of state and transport
coefficients of a hadron gas is necessary for a better understanding of the
observables. Phenomenologically both the transport properties and the equation
of state of hadron gas are the major source of  uncertainties in dissipative
fluid dynamics.

In spite of their importance, the equation of state and transport coefficients
of hot, dense hadron gases are still poorly known because of the
nonperturbative nature of the strong interaction. Progress in the study of
hadron matter transport coefficients is very slow, and only a calculation of
transport coefficients in the variational  method \cite{Prakash,Davesne}
and relaxation time  approximation \cite{Gavin} has been done. From those
previous studies a lot has been learned about the transport coefficients of
binary mixtures such as $\pi\pi$ system. However in a more realistic situation
we need to describe transport properties of a many--body system. This in turn
would require taking into account various interaction processes and in--medium
effects.  Thus, we need to investigate the thermodynamic and transport
properties of a hadron gas by using a microscopic model that includes realistic
interactions among hadrons.  In this work, we adopt a relativistic microscopic
model, UrQMD   and perform molecular--dynamical simulations for a hadron gas of
mesons.

We focus on the hadronic scale temperature ($100$ MeV $< T < 200$ MeV) and zero
baryon number density  which are expected to be realized in the central high
energy nuclear collisions.  Thermodynamic properties and transport coefficients
of hadronic matter in this region should play important roles in dissipative
fluid dynamical models. Sets of statistical ensembles are prepared for the
system of fixed energy density and baryon number density.  Using these
ensembles, the equation of state is investigated.  The statistical ensembles is
then applied in calculating the  shear viscosity coefficient  of a hadron gas
of mesons.

The equation of state of a hot and dense hadron gas had been investigated using
UrQMD \cite{Bass98,Belkacem98}.  The work has provided valuable information
regarding the nature of the hadron gas. In those simulations the temperature
reaches a limiting value with increasing energy density.  This is because in
those simulations  the detailed balance is broken. This in turn leads to the
irreversibility of the equilibrated system.  And without the reversal process of
multi--particle production energy balance between the forward and backward
reactions is no longer realized and hence the saturation of the temperature
occurs.  Although it is interesting and important to formulate these
multi--particle interaction processes exactly in the present simulation,
straightforward implementation of them is not easy. In this work, avoiding this
complicated problem, we disabled three or  many--body interactions in UrQMD. We
have also disabled decays or interactions that involves photons.

The rest of the paper is organized as follows: In section \r{sec:equilibrium}
we study the equilibration and thermodynamics of the system. In section
\r{sec:hadrongas} we study the thermodynamic of a pure resonance meson gas for
comparison with the results from UrQMD. In section \r{sec:viscosity} we
calculate the shear viscosity coefficient from stress tensor fluctuations
around the equilibrium state through UrQMD using Green--Kubo relations.
Finally in section \r{sec:summary} we summarize our results.

\section{Equilibration of infinite matter in a box}
\label{sec:equilibrium}

To investigate the equilibration of the system we performed microscopic
calculation using UrQMD. UrQMD is designed to simulate ultra--relativistic
heavy ion collision experiments.  The description of the model can be found in
\cite{Bass98}.   In studying the equilibration of the hadron gas we would like
to maintain detailed balance in the simulations. Multi--particle productions
plays an important role in the equilibration of the hadron gas. However in
UrQMD their inclusion in the simulations breaks detailed balance due to the
absence of reverse processes. In order to avoid this problem in the present
simulations we consider only up to two--body absorption/annihilation and decay
processes. Thus the fundamental processes in the UrQMD version we use here are
two--body elastic and quasi-elastic collisions between hadrons, and strong
decays of resonances. Even though we started with light mesons in the initial
state we consider  all the mesons and meson--resonance included into the UrQMD
model, in the final state.

When studying the equilibration of hadron gas it is important to maintain
detailed balance in the microscopic model.  Though the contributions of the
multi-particle productions dominate the system at early stages of the
non--equilibrated system, the reverse process plays an important role in the
latter, equilibration stage.  The absence of reverse processes leads to one-way
conversion of the energy to particles.  However, the exact treatment of
multi--particle absorption processes is very difficult. In order to treat them
effectively in our case, we only consider up to 2-body decays. 

In this work, we focus our investigation on the thermodynamic and transport 
properties of a hadronic system.  For this purpose, we consider a
system in a cubic box and impose periodic boundary conditions in
configuration space. Thus if a particle leaves the box, another
one with the same momentum enters from the opposite side. This calculation is
similar to the one done in \cite{Belkacem98} but with different
degree of freedom and included processes in the system. A further similar
analysis was done in \cite{Brat00,Sasaki} using different cascade models with
different degrees of freedom.

The energy density $\varepsilon$ and the baryon number density
$n_B$ in the box are fixed as input parameters, and these
quantities are conserved throughout the simulation.  The initial
distributions of mesons are given by uniform random distributions
in phase space.  
The energy is defined as $\eps=E/V$, where $E$ is the energy of $N$ particles:
\be
E = \sum_{i=1}^{N}\sqrt{m^2_i+p_i^2}
\ee
The 3--momenta $p_{i}$ of the particles in the initial state are randomly
distributed in
the center of mass system of the particles:
\be
\sum_{i=1}^{N} p_{i} = 0 \p
\ee

The time evolution is now described by UrQMD. Though the initial
particles are only $\pi\,,\eta\,,\omega\,,\rho\,,\phi$, 
many mesons and meson--resonances are produced through interactions.
We now propagate all particles in the box using periodic boundary conditions,
that is, particles moving out of the box are reinserted at the opposite side with
the same momentum. The phase--space distribution of mesons then can change
due to elastic collisions, resonance production and their decays to lighter
mesons again. We recall that we include all the mesons and meson--resonances in
UrQMD. 

To investigate the equilibration phenomena of the system we look at the
particle densities and energy distributions of each particle. As time increases
the system tends towards an equilibrium state.  When the system is in thermal
equilibrium, the slope parameters of the energy distributions for all particles
should have the same value, and that value is the inverse of temperature.  To
investigate this, we study the time evolution of the inverse slopes of various
particles.

Running UrQMD many times with the same input parameters and taking the
stationary configuration in equilibrium, we can obtain statistical ensembles
with fixed temperature.  By using these ensembles, we can calculate
thermodynamic quantities, such as the particle density, pressure, and so on, as
functions of temperature and baryon number density.  We extract the shear
viscosity coefficient by finding the energy--momentum  tensor correlations and
then employ the Green--Kubo relations..

We specify the initial input parameters: the volume of the box $V$, 
the net baryon number density $n_B$, and the
total energy density $\eps$. 
We consider the input parameters which will give the temperature
range 100 -- 200 MeV.  Here $n_{B} = 0.0$ fm$^{-3}$ is taken as the 
net baryon number density of the system. 
We generated a statistical ensemble of 200 events.

\subsection{Chemical Equilibration}

\begin{figure}[h]
\centerline{\psfig{figure=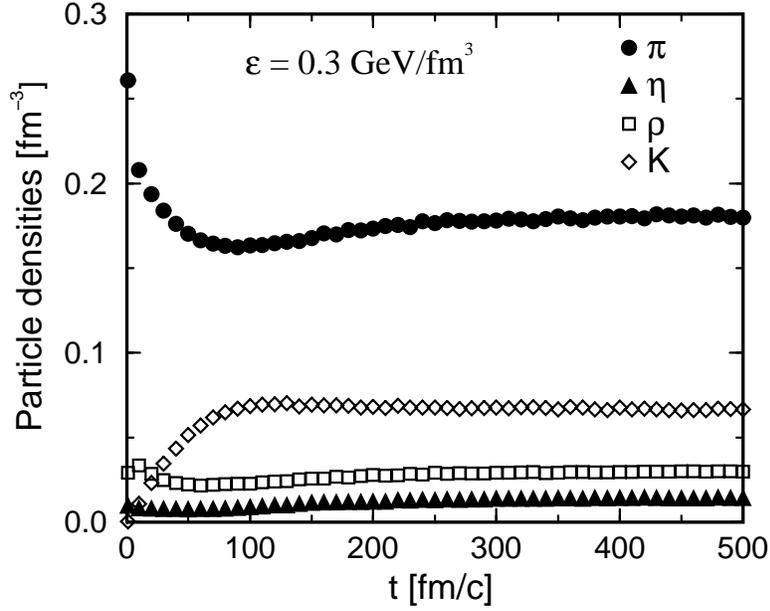,width=11cm}}
\caption{\label{fig:ndens03}
The time evolution of particle densities for each particle with
    $V = 1000$ fm$^{3}$ and $\varepsilon=0.3$ GeV/fm$^{3}$.
}
\end{figure}

\begin{figure}[h]
\centerline{\psfig{figure=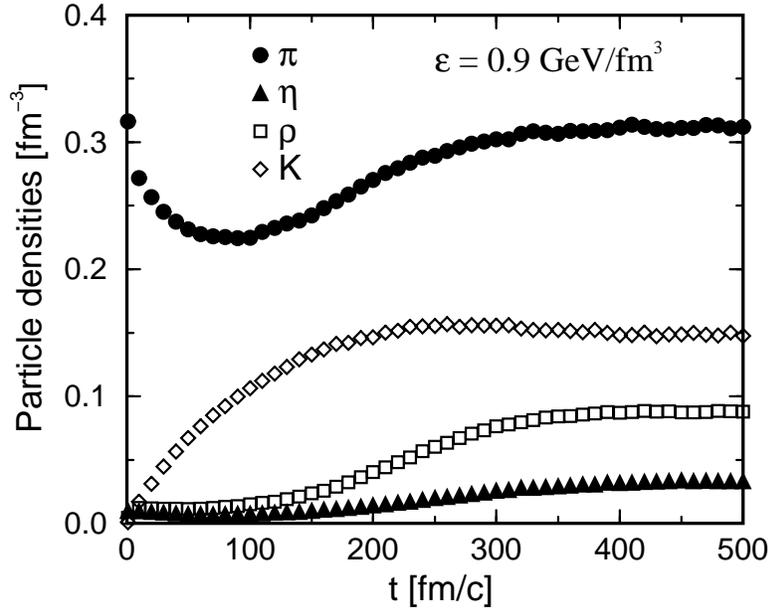,width=11cm}}
\caption{\label{fig:ndens09}
The time evolution of particle densities for each particle with
    $V = 1000$ fm$^{3}$ and $\varepsilon=0.9$ GeV/fm$^{3}$
}
\end{figure}

Figure \ref{fig:ndens03} shows the time evolution of the various particles
densities ($\pi,\eta,\rho,K$) at zero net baryon number density and energy
density $\eps = 0.3$ GeV/fm$^3$. After several fm/c the number of pions
decreases first due to inelastic collisions and annihilation that produces other
meson resonances. The pion density then increases due to decay of heavier meson
resonances to an equilibration. The number of kaons (in general strange
mesons) increases to equilibration value in much longer times than other
particles. 
In figure \ref{fig:ndens09} we show the same situation but with different
initial energy density of the box, $\eps = 0.9$ GeV/fm$^3$. For large initial
energy densities the equilibration times are much larger.

Figures \ref{fig:ndens03} and \ref{fig:ndens09} display the time evolution
of particle densities.  These figures show that the system
approaches a stationary state with time.  The saturation of
particle densities indicates the realization of chemical
equilibrium. We conclude that
chemical equilibrium in our system is realized.

\subsection{Thermal Equilibration and Temperature}

In this subsection we investigate the approach to thermal equilibrium. This is
driven by the momentum equilibration of the system. That is, when the momentum
anisotropy of the system has dropped to a limiting value such that the system can
be described by simple global thermodynamic variables like temperature. The
thermal equilibration times have to be contrasted to those for chemical
equilibrium.

\begin{figure}[hp]
\begin{minipage}[t]{7.0cm}
\centerline{\psfig{figure=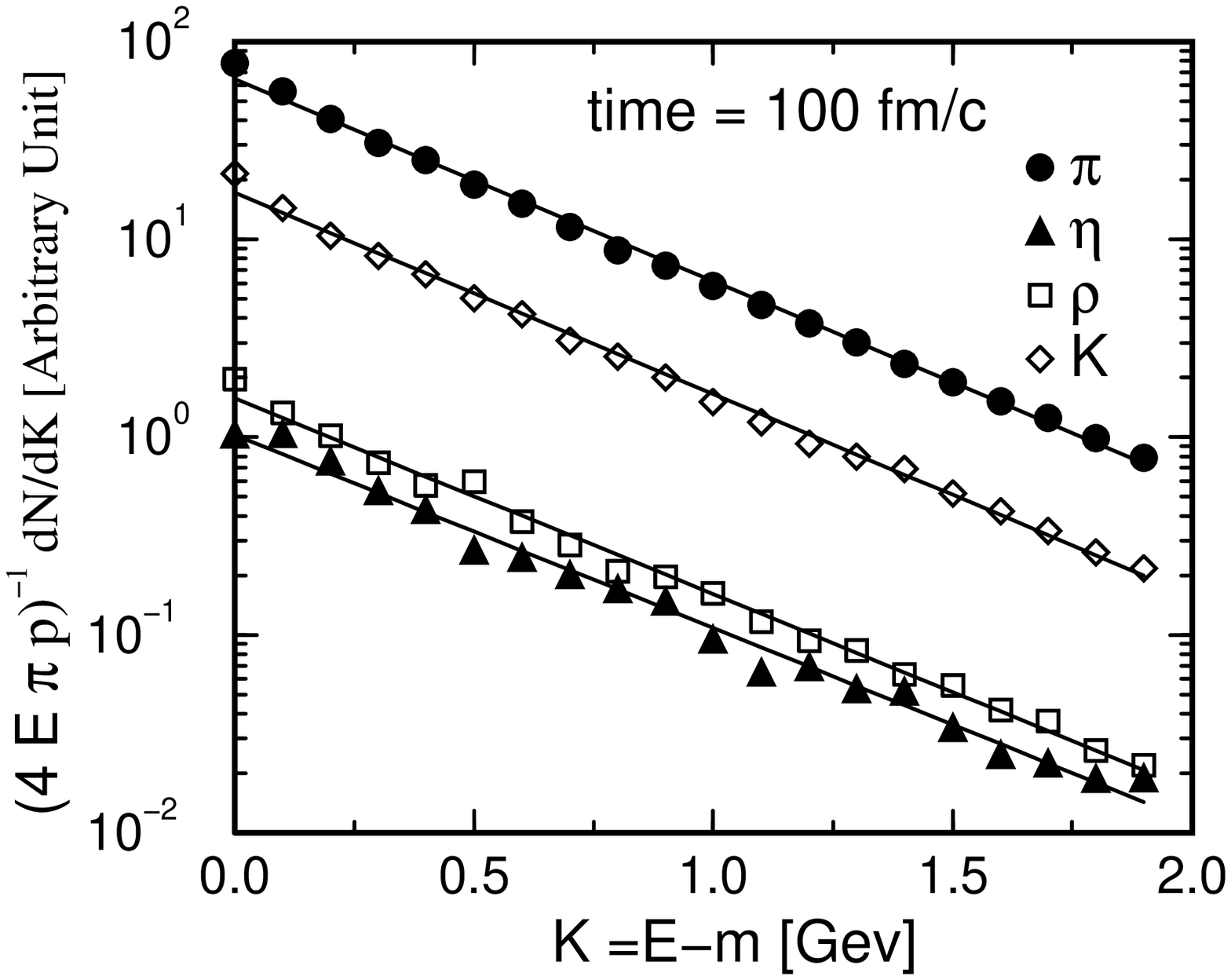,width=7.5cm}}
\end{minipage}
\hfill
\begin{minipage}[t]{7.0cm}
\centerline{\psfig{figure=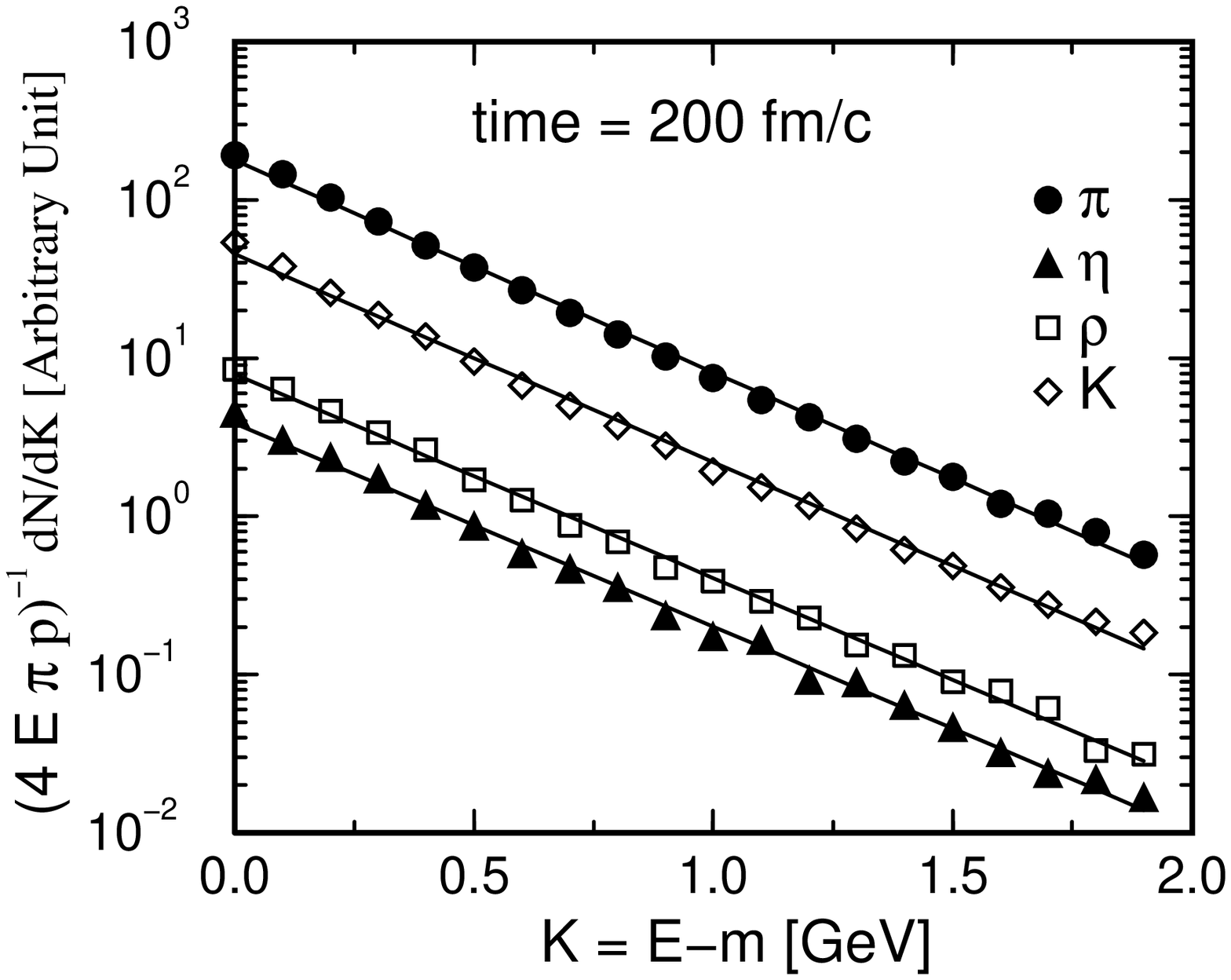,width=7.5cm}}
\end{minipage}

\vspace{0.3cm}

\begin{minipage}[b]{7.0cm}
\centerline{\psfig{figure=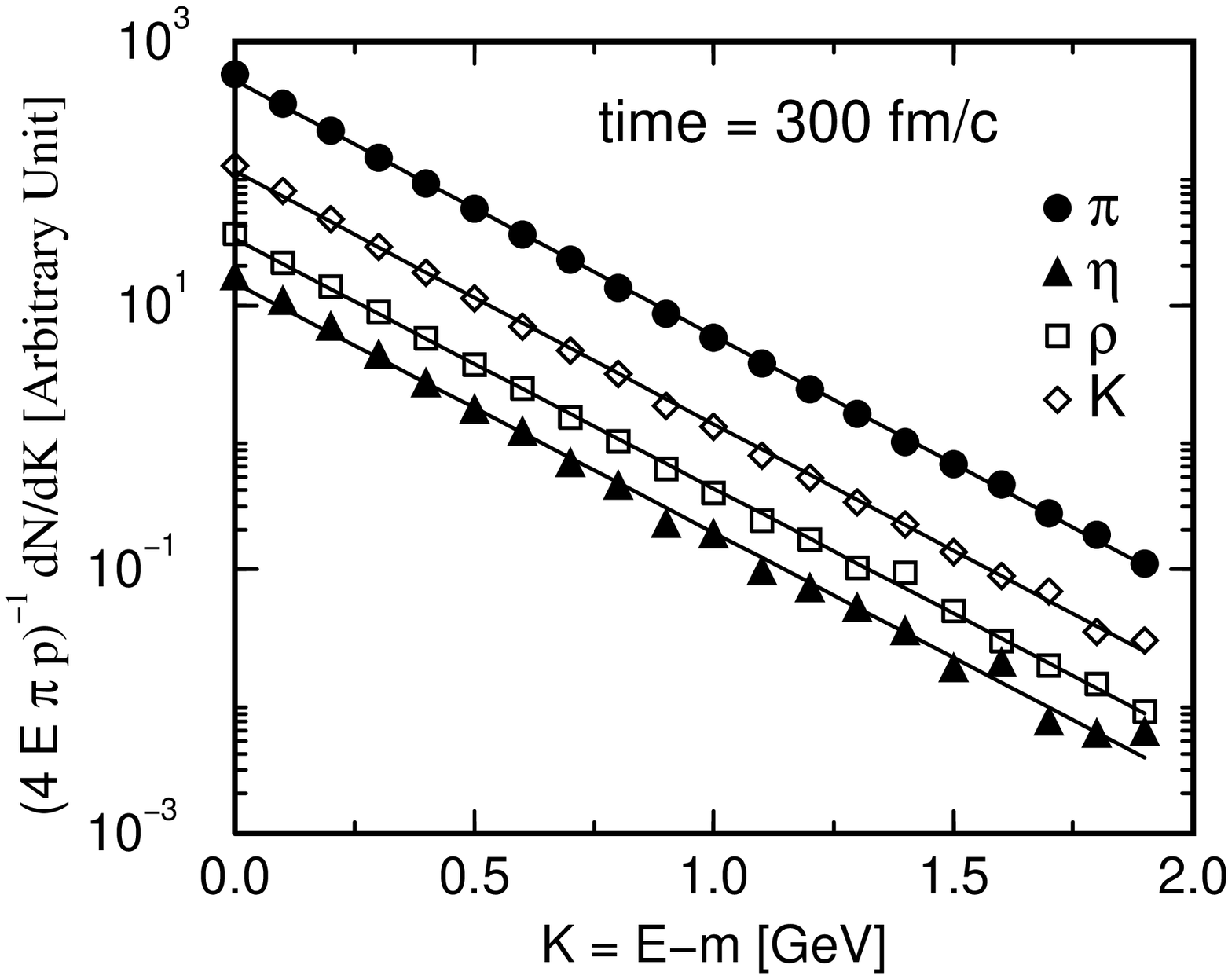,width=7.5cm}}
\end{minipage}
\hfill
\begin{minipage}[b]{7.0cm}
\centerline{\psfig{figure=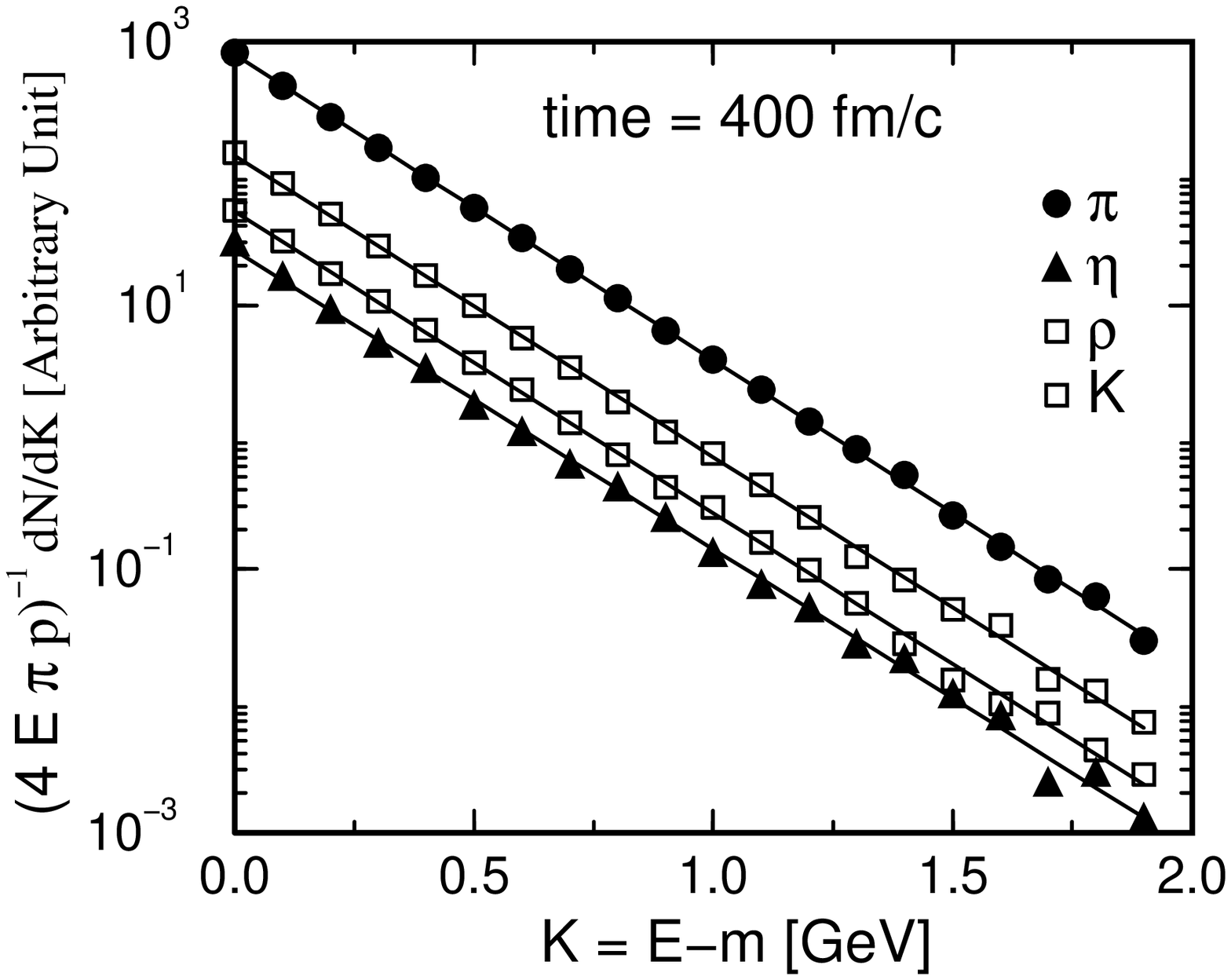,width=7.5cm}}
\end{minipage}
\caption{\label{fig:distribution}Energy distributions of $\pi$, $\eta$, $\rho$ and $K$ 
    at four different values of time, $t=100$  fm/c,
    $t=200$ fm/c, $t=300$ fm/c and $t=400$ fm/c.  The lines are the
    fitted results that are given by Boltzmann distributions,
    $C\exp(-\beta E)$.  The calculation was done with
    $V=1000$ fm$^{3}$, $n_{B}=0.0$ fm$^{-3}$ and
    $\varepsilon=0.9$ GeV/fm$^{3}$. }
\end{figure}

Figure \ref{fig:distribution} displays energy distributions of $\pi,
\eta, \rho$ and $K$ at time $t=100,\;200,\;300$ and $400$ 
fm/c. For equilibrated system the energy distributions approach the Boltzmann
distribution,
\begin{equation}
  \frac{dN_i}{d^{3} p} = \frac{dN}{4\pi Ep dE}
  = C\exp(-\beta E_i),
  \label{eq:Boltz}
\end{equation}
as time increases, where $\beta$ is the slope parameter of the
distribution. Here $E_i =(p_i^2+m_i^2)^{1/2}$ is the energy of particle $i$. 
Moreover, the slopes of the energy distributions
converge to a common value.  These results indicate 
realization of thermal equilibrium. 

\begin{figure}[hp]
\begin{minipage}[t]{7.0cm}
\centerline{\psfig{figure=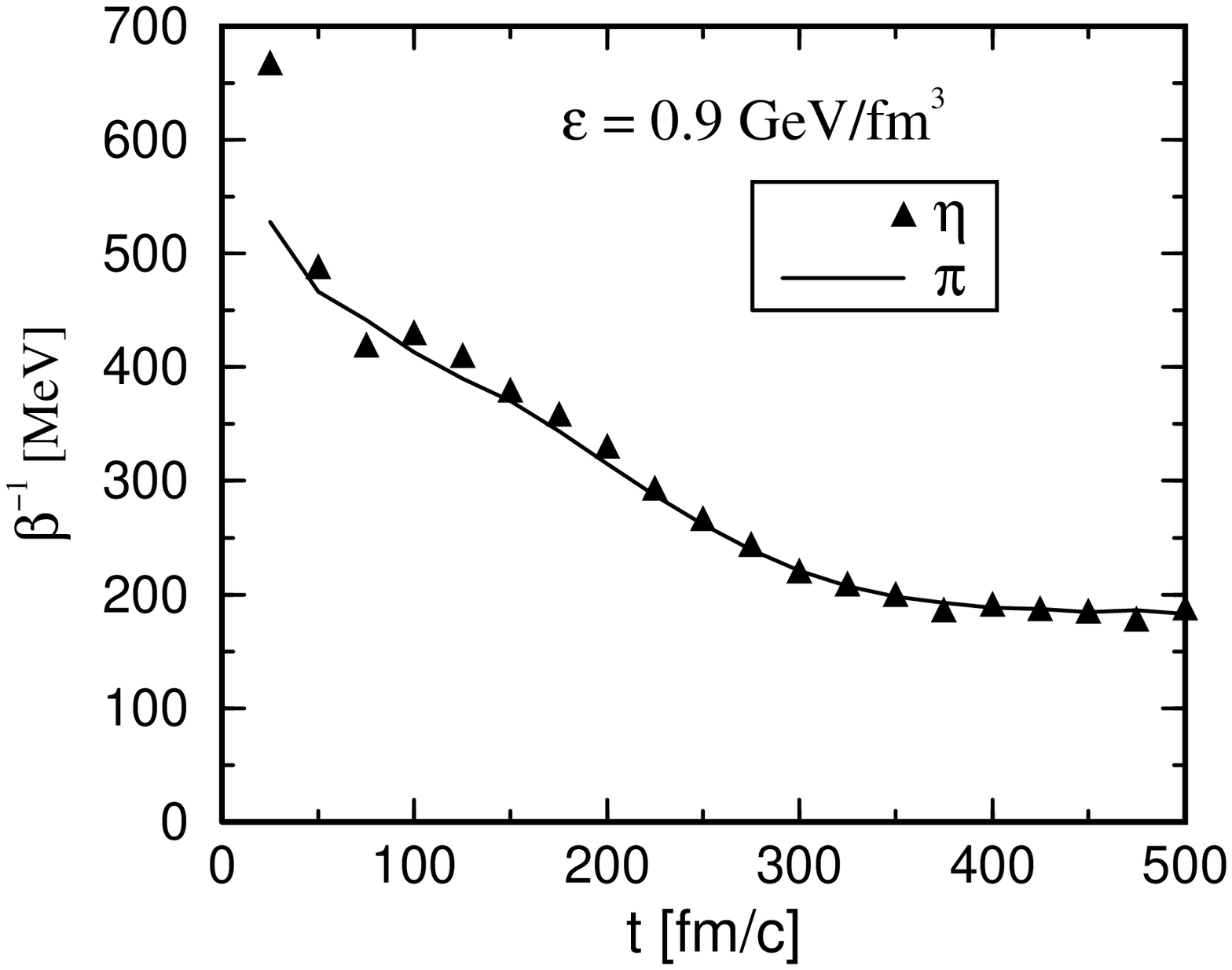,width=7.5cm}}
\end{minipage}
\hfill
\begin{minipage}[t]{7.0cm}
\centerline{\psfig{figure=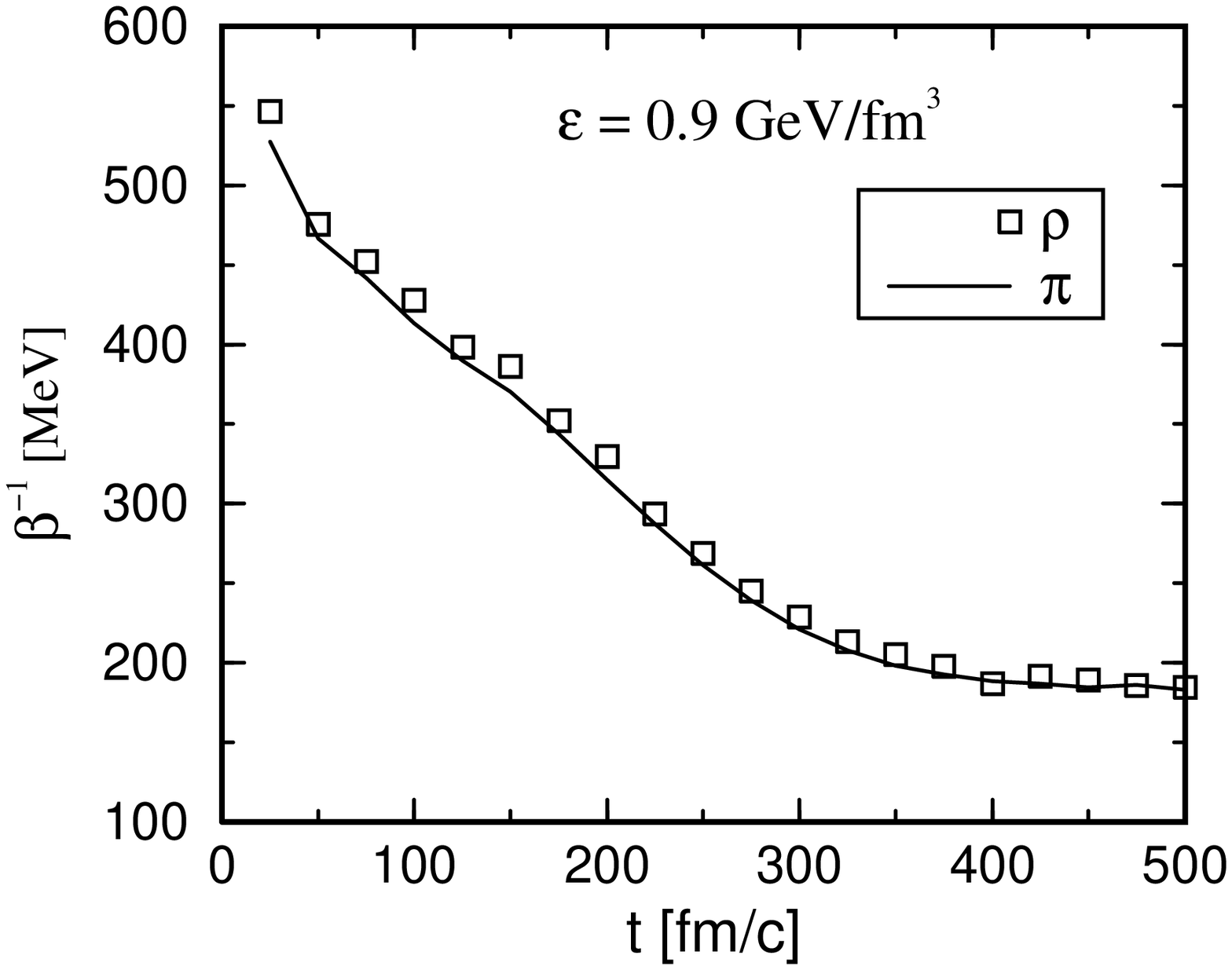,width=7.5cm}}
\end{minipage}

\vspace{0.3cm}
\centering
\begin{minipage}[b]{9cm}
\centerline{\psfig{figure=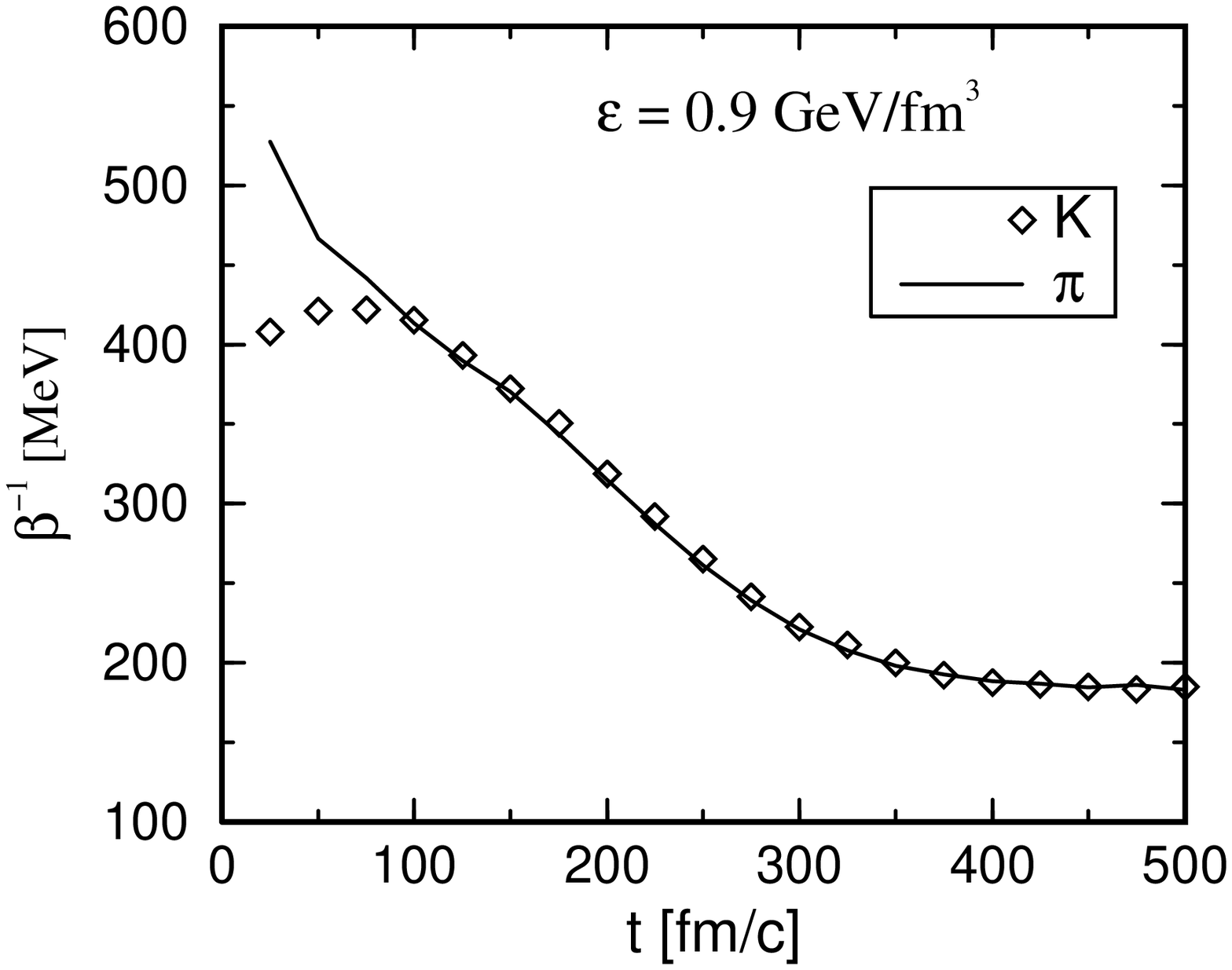,width=9cm}}
\end{minipage}

\caption{\label{fig:slope}The time evolution of the inverse slopes $\beta^{-1}$ for
    $\pi$, $\eta$, $\rho$ and $K$ with
    $V=1000$ fm$^{3}$, $n_{B}=0.0$ fm$^{-3}$,
    $\varepsilon=0.9$ GeV/fm$^{3}$.  The value of $\beta^{-1}$ was
    calculated from the fitting of energy distributions. Here the
    solid curves represent the time evolution of $\beta^{-1}$ for
    $\pi$.}
\end{figure}

Figure \ref{fig:slope} displays the time evolution of the
inverse slopes of different particle species that were calculated 
by fitting the energy
distributions to a Boltzmann distribution. The solid
curves correspond to the time evolution of the inverse slope of pions.  
From this figure, it
is seen that the difference between the pion inverse slope and other
particles' inverse slopes become zero for times latter than $350$ fm/c.
Therefore, we conclude that thermal equilibrium is established
at about $t = 350$ fm/c; the values of the inverse slope
parameters of the energy distribution for all particles become
equal for latter times.  Thus we can regard this value as the
temperature of the system. 
The equilibration time is large. If we allow for multi--particle production
and absorption the equilibration time would be shorten significantly. 

\section{Hadronic gas model}
\label{sec:hadrongas}

In this subsection we compare the UrQMD box calculations with a simple
statistical model for an ideal hadron gas where the system is described by a
grand canonical ensemble of noninteracting bosons in equilibrium at temperature
$T$. All meson species considered in UrQMD are also been used in the statistical
model. In hadron gas model we use as input the same energy density and net
baryon density to obtain the temperature of the system.

In hadron gas we find that the temperature increases continuously with energy
density.

\begin{figure}[hp]
\centerline{\psfig{figure=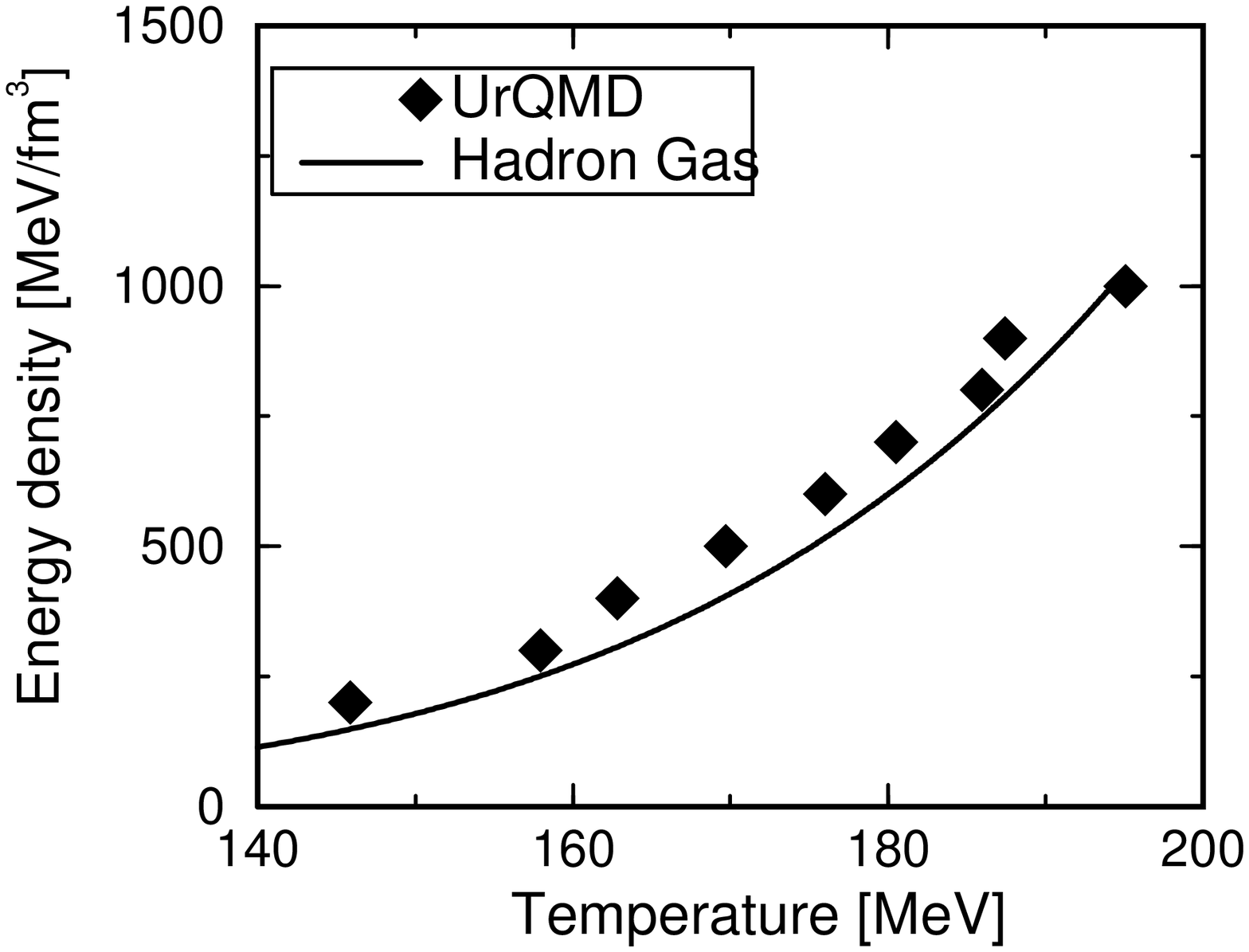,width=9cm}}
\caption{\label{fig:epstemp}The equation of state of a mixed hadron gas at finite temperature
    ($100$ MeV $< T < 200$ MeV) and zero baryon density
    (0.0 fm$^{-3}$).  The energy
    density of mesons is plotted as functions
    of the temperature. The curve corresponds to the free gas model 
    represented by Eq. (\ref{eq:epstemp}).
 }
\vspace{0.3cm}
\centerline{\psfig{figure=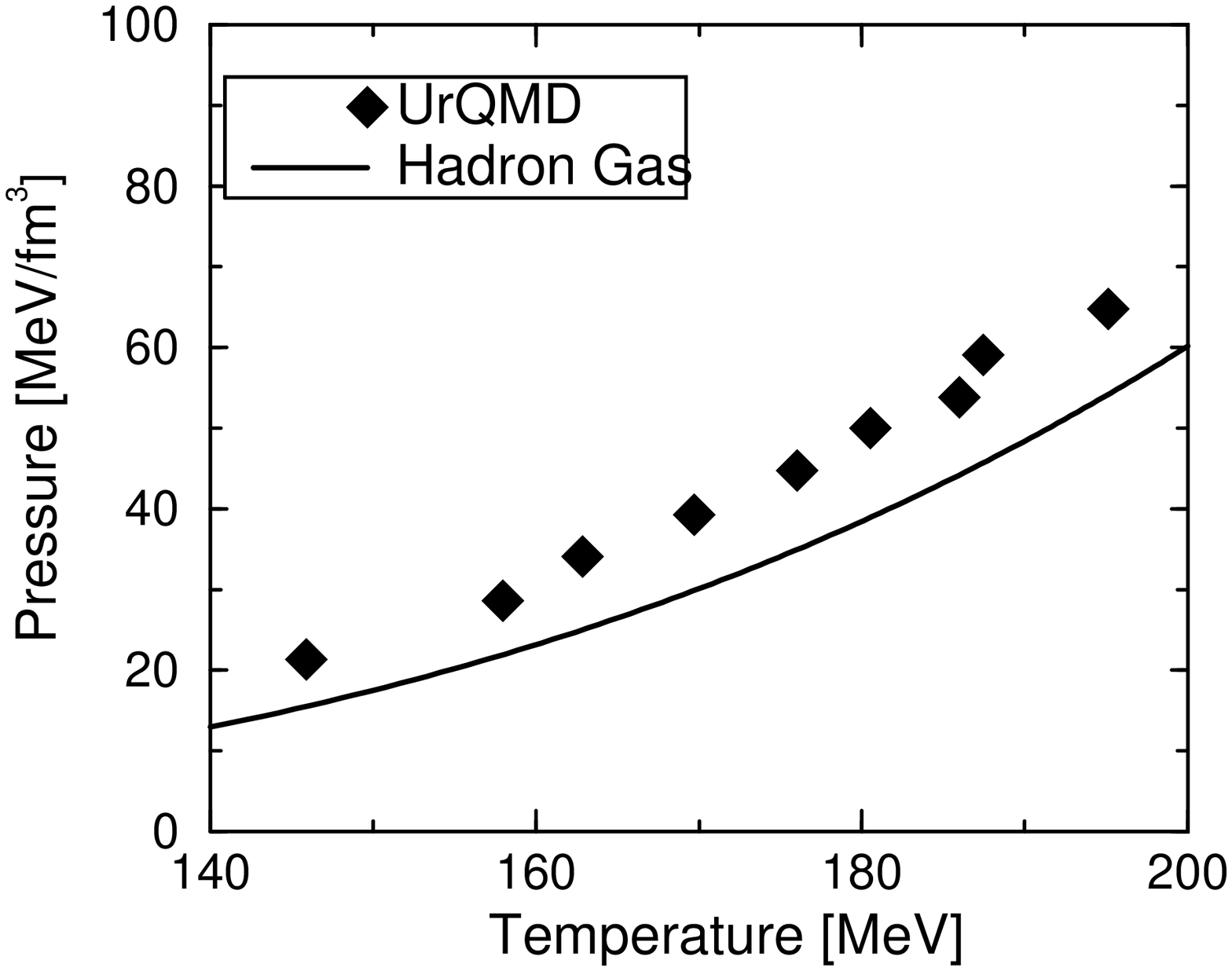,width=9cm}}
\caption{\label{fig:presstemp}The equation of state of a mixed hadron gas at finite temperature
    ($100$ MeV $< T < 200$ MeV) and zero baryon density
    (0.0 fm$^{-3}$).  The pressure 
    of pions is plotted as functions
    of the temperature. The curve corresponds to the free gas model 
    represented by Eq. (\ref{eq:presstemp}).
 }
\end{figure}

Figures \ref{fig:epstemp} and \ref{fig:presstemp} show the relations between 
the temperature and thermodynamic quantities such as energy
density, 
\be
 \eps = \frac{1}{V}\sum_{i=1}^{\mbox{\scriptsize all particles}}E_{i} \K
 \ee 
 particle density, and pressure,
\be
 P = \frac{1}{3V}\sum_{i=1}^{\mbox{\scriptsize all particles}}
    \frac{{p}_{i}^2}{E_{i}} \p
\ee  
In these figures, all curves correspond to the relativistic Bose-Einstein gas 
\begin{eqnarray}
\label{eq:eos}
  \label{eq:eoset}
  \varepsilon(T,\mu) & = &
  \sum_{k} g_{k}\int\frac{d^{3}p}{(2\pi)^{3}}
  \frac{ E_k}{e^{\frac{E_k-\mu}{T}} - 1},\\
  \label{eq:epstemp}
  n(T,\mu) & = &
  \sum_{k} g_{k}\int\frac{d^{3}p}{(2\pi)^{3}}
  \frac{1}{e^{\frac{E_k-\mu}{T}} - 1},\\
  \label{eq:ntemp}
  p(T,\mu) & = &
  \sum_{k} g_{k}\int\frac{d^{3}p}{(2\pi)^{3}}
  \frac{p^{2}}{3E_k}
  \frac{1}{e^{\frac{E_k-\mu}{T}} - 1}, \l{eq:presstemp} 
\end{eqnarray}
where $g_{k}$ is a degeneracy factor. In these calculations  
the meson chemical potential $\mu$ is fixed to zero.

Figure \ref{fig:epstemp}  shows the energy density versus temperature for 
mesons. In this figure,   the difference between UrQMD results and those for
the calculation of the free gas model is negligible.  Figure
\ref{fig:presstemp} shows the pressure versus temperature for mesons.   There
is deviation of UrQMD results from the free gas model results especially at
high temperatures. The influence of interactions is clear above $T \sim m_\pi$.
Enhancement of heavy meson resonances grows as the temperature  increases.

In a previous study \cite{Belkacem98} \ the limiting value of temperature with
increasing energy appeared. As already mentioned this is because of the lack of
reversal process of multi--particle production in that study. In this
calculation where we try to maintain detailed balanced in UrQMD, this limiting
temperature does not appear. This is an important result of
taking detailed balance into account.

However, in this simulation the lack of multi--particle production leads to
long equilibration times. This is also because we do not have  meson--baryon
interactions, such as $\pi N \rightarrow R$ and their inverse processes.  The
enhancement of heavy baryon resonances causes an increase in the abundances of
mesons, and vice versa. Heavy resonances readily produce two pions, and thus
the enhancement of heavy baryon resonances promotes meson production. 
Therefore, interactions between mesons and baryons are very important in the
study of the properties of a mixed hadron gas. Inclusion of multi--particle
interactions would also shorten the equilibration time considerably.

\section{Shear Viscosity Coefficient}
\label{sec:viscosity}

Transport coefficients such as viscosities, diffusivities and conductivities
characterizes the dynamics of fluctuations of dissipative fluxes in a medium.
Transport coefficients can be measured, as in the case of condensed matter
applications. However in principle they should be calculable theoretically from
first principles.

In a weakly coupled theory transport coefficients can be computed in a 
perturbative expansion, employing
either kinetic theory or field theory using Kubo formulas
\cite{Baym,Jeon,Arnold,Hosoya,Heiselberg,Aarts,Hou}. The resulting Kubo relations \cite{Kubo} 
express transport coefficients in terms of the zero-frequency slope of 
spectral densities of current-current, or stress tensor-stress tensor 
correlation functions,

Monte Carlo simulations for transport coefficients is a powerful tool when
studying transport coefficients using Green--Kubo relations. For calculation of
transport coefficients of shear viscosity, thermal conductivity, thermal
diffusion and mutual diffusion for a binary mixture of hard spheres see
\cite{Erpenbeck} and for the calculation of diffusion coefficient of a hadron gas
see \cite{Sasaki}

Knowledge of various transport coefficients is important in dissipative fluid
dynamical models \cite{Muronga03}. In this paper we consider the evaluation
of shear viscosity coefficient of a hadron gas of mesons and their resonances.

In trying to stay close to the extended irreversible thermodynamic processes we 
will, however, use the
Kubo formulas in fluctuation theory to extract transport coefficients.

In the longitudinal boost--invariant flow the
important coefficient is the shear viscosity. 
In dissipative fluids the expression for the entropy 4--current is governed by
transport coefficients and relaxation coefficients. These coefficients determine the strength of the
fluctuations of dissipative fluxes about the equilibrium state. The  
generalized entropy plays an important role in the description of the
fluctuations of conserved quantities and of the dissipative fluxes.

\begin{figure}[ph]
\begin{minipage}[t]{7.0cm}
\centerline{\psfig{figure=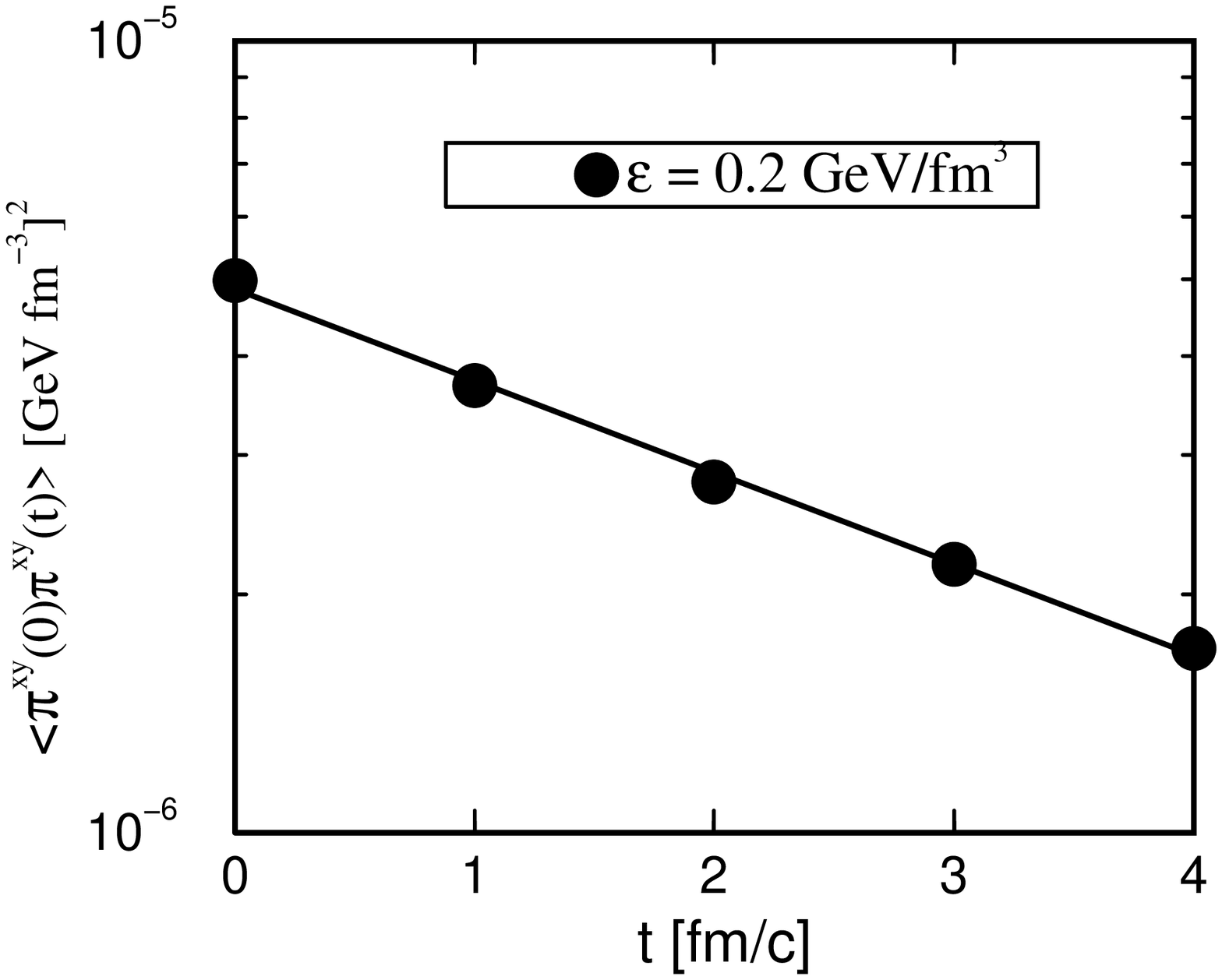,width=7.5cm}}
\end{minipage}
\hfill
\begin{minipage}[t]{7.0cm}
\centerline{\psfig{figure=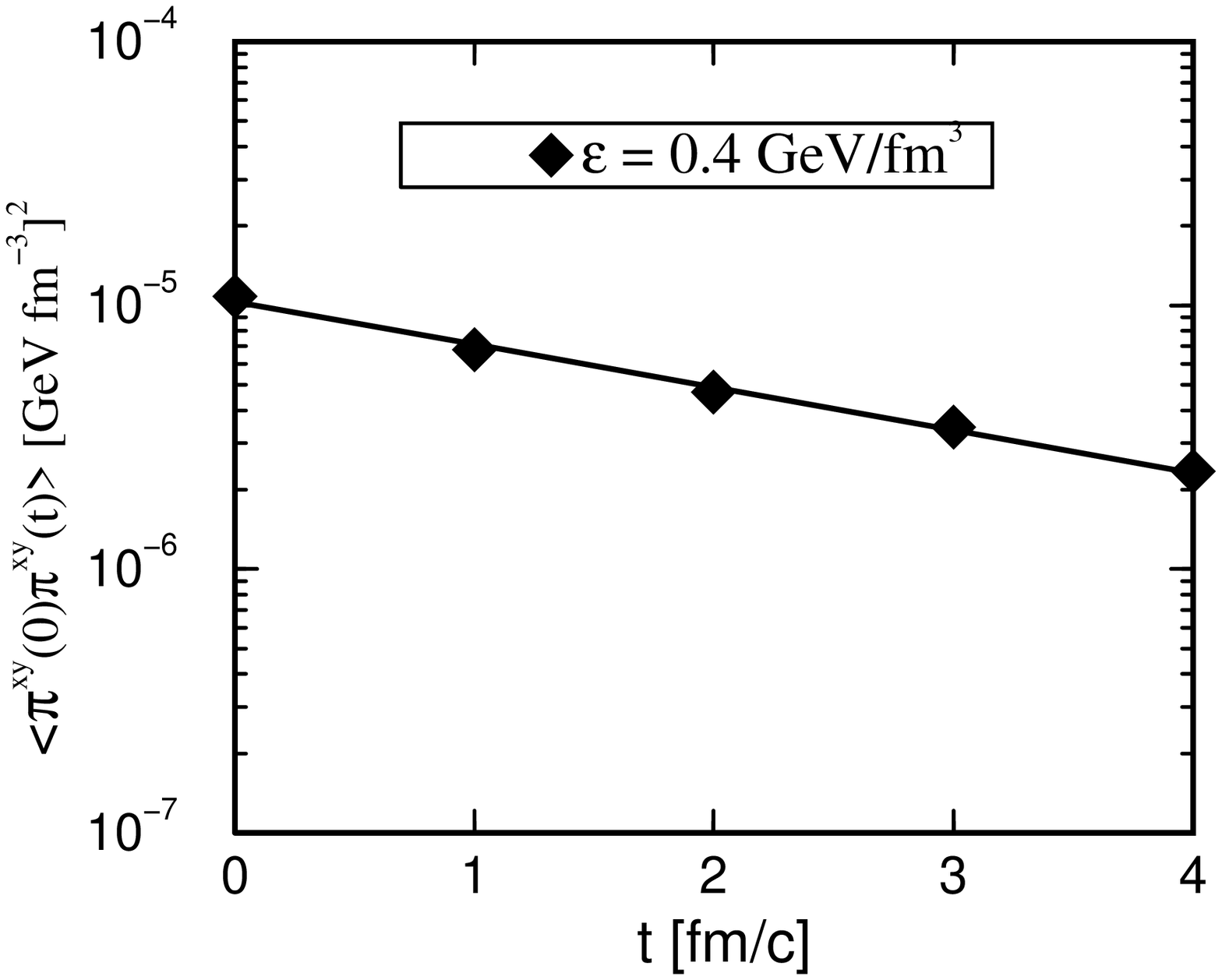,width=7.5cm}}
\end{minipage}

\vspace{0.3cm}

\begin{minipage}[b]{7.0cm}
\centerline{\psfig{figure=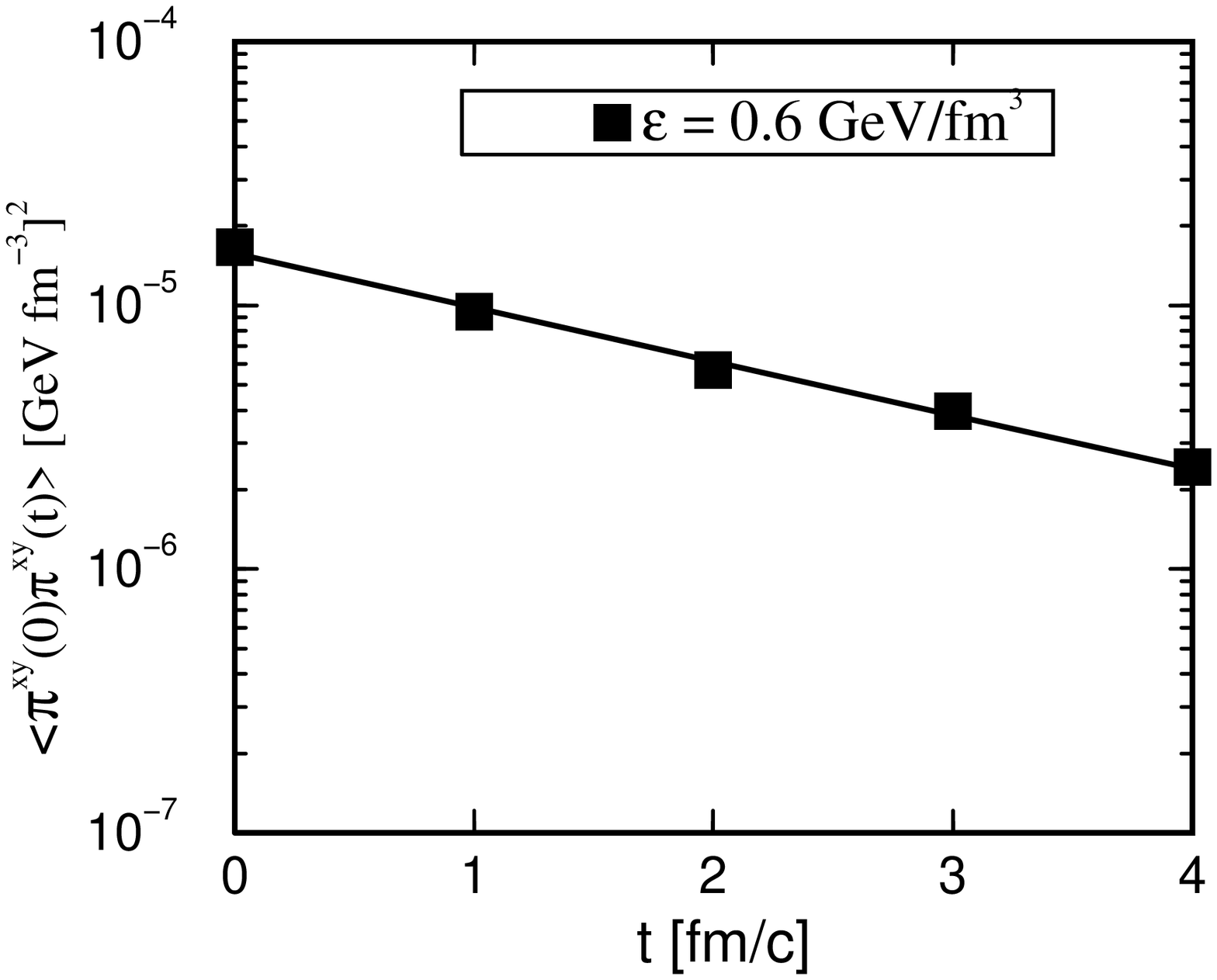,width=7.5cm}}
\end{minipage}
\hfill
\begin{minipage}[b]{7.0cm}
\centerline{\psfig{figure=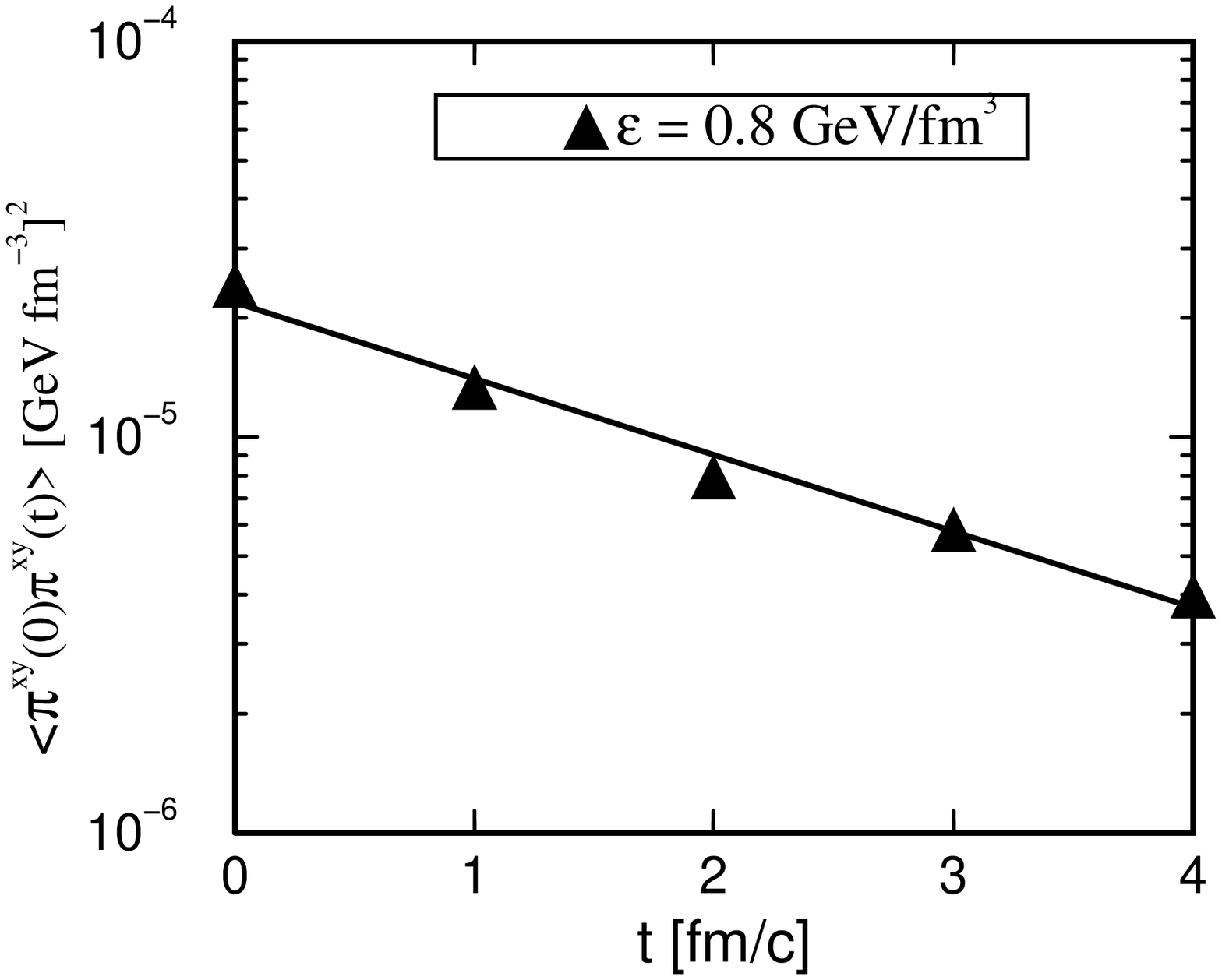,width=7.5cm}}
\end{minipage}
\caption{\label{fig:pipicor} 
Stress--tensor correlation of the mesons as a function of time. The curves are
the exponential fits to extract relaxation times}
\end{figure}

Now we calculate the coefficient of shear viscosity. 
First, the fluctuation--dissipation
theorem tells us that shear viscosity $\eta$ is given by the
stress tensor correlations \cite{Kubo}
\begin{equation}
  \eta ={V\over T}
  \int_{0}^{\infty}\lla\pi_{ij}(t)\cdot \pi_{ij}(t+t')
  \gra dt'\K
  \label{eqn;fdt}
\end{equation}
where $\pi_{ij}\equiv T_{ij} - \delta_{ij}{\mathcal {P}}$  
denotes the traceless part of the stress tensor 
and ${\mathcal{ P}} \equiv {1\over3} T^{i}_{\;\;i}$ the (local) pressure. 
The angular brackets stand for equilibrium average, i.e., average over the
number of ensemble states and average over the number of particles.
The correlation functions are damped exponentially with time
(see Fig. \ref{fig:pipicor}):
\begin{equation}
  \lla\pi_{ij}(t)\cdot \pi_{ij}(t+t')\gra \propto
  \exp{\left (- \frac{t'}{\tau_{\pi}}\right )}.
  \label{eqn;rlx}
\end{equation}
The solid lines in Fig. \r{fig:pipicor} are the fits to the correlations and the
inverse slope corresponds to the relaxation time. 
The shear viscosity coefficient can be rewritten in the simple form
\begin{equation}
  \eta = {V\over T}
  \lla\pi_{ij}(t)\cdot \pi_{ij}(t)\gra \tau_{\pi},
  \label{eqn;difcon}
\end{equation}
where $\tau_{\pi}$ is the relaxation time of the shear flux.
In this work we used a box of volume $V = 1000$ fm$^{3}$. The results are
insensitive to the box length greater than 6 fm.

To this end, we have to remark that the transport coefficients represents the
fluctuations of the dissipative fluxes around an equilibrium state. In terms of
fluctuations the Green-Kubo relation (at zero frequency) for 
shear viscosity can be written as 
\be
\eta =  {V\over T} \int_0^\infty \lla\delta \pi_{ij}(0)\delta \pi_{ij}(t)\gra dt
~~~~~(i\neq j) \K \label{eq·shearvis}
\ee
In the above equation the 
fluctuations of shear flux are exponentially damped. They are 
obtained found from 
the second differential of the generalized entropy expression \cite{Muronga03}
\be
\lla\delta \pi_{ij}(0) \delta \pi_{kl}(t)\gra = \eta T (\tau_\pi V)^{-1}
\btu_{ijkl} \exp(-t/\tau_\pi)\p \label{eq:2ndmom}
\ee
with  
$\btu_{ijkl} = (\delta_{ik}\delta_{jl}+\delta_{il}\delta_{jk}
- (2/3)\delta_{ij}\delta_{kl})$. In the limit of vanishing relaxation times, we
recover the formulae of Landau and Lifshitz, since in this limit $\tau^{-1}
\exp(t/\tau) \rightarrow 2\delta(t)$ with $\delta(t)$ the Dirac delta function.
Equation (\r{eq:2ndmom}) relates the dissipative coefficient 
$\eta$ to the fluctuations of the fluxes with respect to
equilibrium. We see that fluctuations determine the dissipative coefficients.
Conversely, transport coefficients determine the strength of the fluctuations.

If the evolution of the fluctuations on the fluxes is described by the
Maxwell--Cattaneo (see \cite{Muronga03}) relation equations then after integration the above
expression for the shear viscosity coefficient reduces to
\be
\eta = {\tau_\pi V\over T}  \lla\delta \pi_{ij}(0)\delta \pi_{ij}(0)\gra \p
\ee
In what follows we will use the existing microscopic model, namely
UrQMD, to extract the shear viscosity coefficient.

\begin{figure}[htb]
\centerline{\psfig{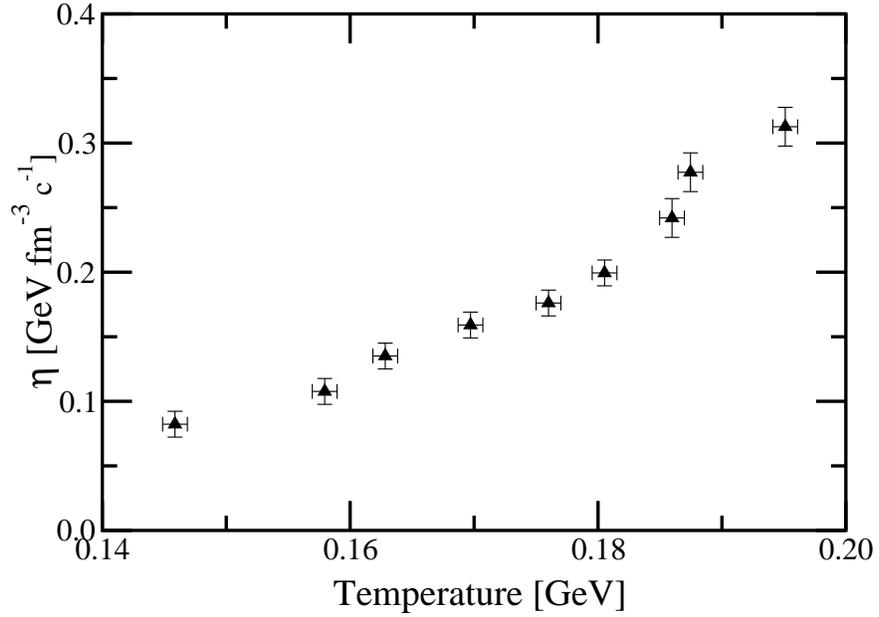}}
\caption{\label{fig:urqmdshear}Shear viscosity of meson gas as a function of
temperature.}
\end{figure}

\begin{figure}[hbt]
\centerline{\psfig{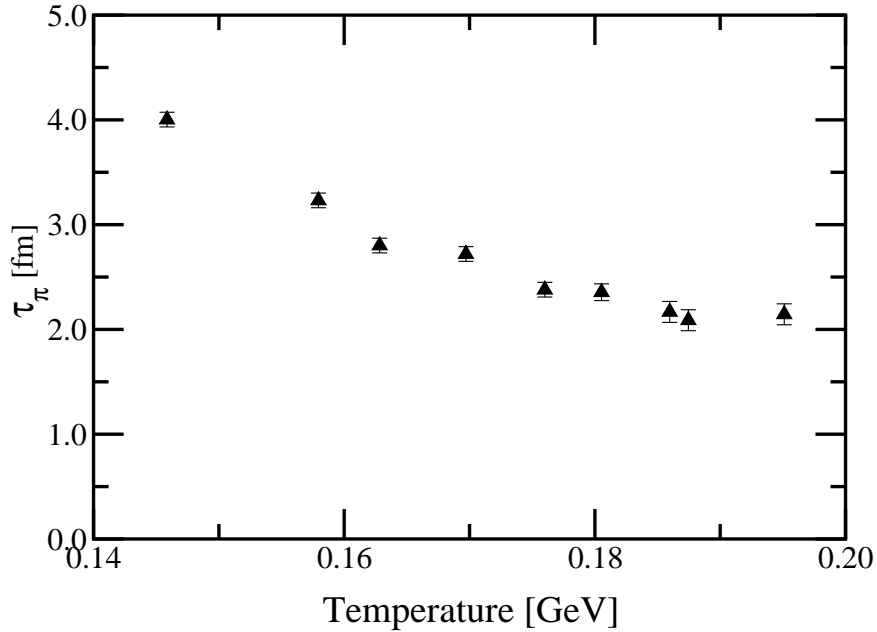}}
\caption{\label{fig:urqmdrelax}The relaxation time for the shear flux of meson 
gas as a function of temperature.}
\end{figure}

Figure \r{fig:urqmdshear} shows the shear viscosity coefficient results from
UrQMD using Kubo relations. As in the variational approach the coefficient
grows with temperature. The UrQMD results are about twice those from the
variational method. This might be due to the many meson resonances included in
UrQMD while  in the  variational method we only have pions. Also the cross
section parameterizations are different in the two approaches.  Figure
\r{fig:urqmdrelax} shows the relaxation time for shear flux in a hot pion gas
calculated from UrQMD by fitting the shear stress correlations. The dependence
of the shear relaxation time on temperature is similar to the one obtained
using variational method. The results obtained here are about a factor of two
less than variational method results. The reasons are similar to the ones given
above for the shear viscosity coefficient.

\section{Conclusions and outlook}
\label{sec:summary}
The transport coefficients for a hadron gas can be obtained easily from
microscopic transport models such as UrQMD. The study of fluctuations of
dissipative fluxes around equilibrium yields Green--Kubo relations which are
more easily applied. The use of fluctuations through Kubo relations has the
advantage of finding not only the transport coefficients but also the
corresponding relaxation times. In addition it is also possible to obtain the
relaxation coefficients such as $\beta_2$ used in \cite{Muronga03}. 

Since the shear viscosity coefficient for QCD has been calculated by many
authors using either kinetic theory or pertubative expansion, it will be
interesting to calculate the shear viscosity coefficient for quark gluon plasma
using microscopic models in the form of parton cascade models such as VNI/BMS
\cite{BMS}. This is currently under investigation \cite{Muronga04}.

\acknowledgments
I would like to thank Joe Kapusta and Horst St\"ocker for valuable comments.
This work was supported by the US Department of Energy grant DE-FG02-87ER40382.

\end{document}